\input harvmac

\def \ha {\half}
\def \ov {\over}

\def \four{{\textstyle {1\ov 4}}}
\def \a {\alpha}
\def \lr { \lref}

\def \ha{{\textstyle{1\over 2}}}

\def \a {\alpha}
\def \b {\beta}
\def \zeta {\zeta}

\def \t {\theta}
\def \td {\tilde }

\def \ov {\over }
\def \four{{\textstyle{1\over 4}}}

\def   \td {\tilde }

\def\({\left( }
\def\){\right)}
\def\N{ {\cal N} }
\def\sei{ {\textstyle{1\ov 6} } }
\def\ads {$AdS_7$}
\def\adss{$AdS_7 \times S^4$}

\def \lr { \lref}
\def\np {{  Nucl. Phys. }}

\lref\thooft{G. 't Hooft, ``A Planar Diagram Theory for Strong Interactions,'' Nucl. Phys. {\bf B72} (1974) 461.}

\lref\polya{A. M. Polyakov, ``String Theory and Quark Confinement,''
hep-th/9711002.}

\lref\malda{J. Maldacena,  ``The Large
$N$ Limit of Superconformal Field Theories and Supergravity,''
hep-th/9711200.} 

\lref\ewt{E. Witten, ``Some Comments on String Dynamics,'' in {\it Strings '95},
ed. I. Bars et. al. (World Scientific, 1997), hep-th/9507121;
N.~Seiberg and E.~Witten, \np B471 (1996) 121.
}

\lr\hoog{G. T. Horowitz and H. Ooguri, ``Spectrum of Large $N$ Gauge Theory
from Supergravity,'' hep-th/9802116.}

\lref\oldstr{A. Strominger, ``Open $p$-Branes,'' Phys. Lett. {\bf B383} (1996) 44, hep-th/9512059.} 

\lref\duff{M. J. Duff, B. E. W. Nilsson,
and C. N. Pope, ``Kaluza-Klein Supergravity,'' Physics Reports
{\bf 130} (1986) 1l.}

\lref\oduff{M. J. Duff, H. Lu,
and C. N. Pope, ``$AdS_5\times S^5$ Untwisted,'' hep-th/9803061.} 

\lr\kach{S. Kachru and E. Silverstein, ``4d Conformal Field Theories
and Strings on Orbifolds,'' hep-th/9802183.}

\lr\vaffa{M. Bershadsky, Z. Kakushadze, and C. Vafa,
``String Expansion as Large $N$ Expansion of Gauge Theories", hep-th/9803076.}

\lr\horro{G. T. Horowitz and S. F. Ross, ``Possible Resolution
of
Black Hole Singularities from Large $N$ Gauge Theory,'' hep-th/9803085.}
\lr\edw{E. Witten,  ``Anti-de Sitter Space, Thermal Phase Transition, and Confinement in Gauge Theories'',
hep-th/9803131.}

\lr\ity{N. Itzhaki, A. A. Tseytlin and S. Yankielowicz, ``Supergravity Solutions for Branes Localized within Branes",
hep-th/9803103.}

\lr\wittn{E. Witten,  ``Anti-de Sitter Space and Holography,''
hep-th/9802150.}

\lref\melvin{J.G.~Russo and A.A.~Tseytlin, ``Magnetic Flux Tube Models
in Superstring Theory", \np B461 (1996) 131.}

\lref\mastro{J.~Maldacena and A.~Strominger, ``$AdS_3$
Black Holes and a Stringy Exclusion Principle",
hep-th/9804085.}

\lr\gkp{S. S. Gubser, I. R. Klebanov, and A. M. Polyakov, ``Gauge Theory Correlators from Noncritical String Theory,'' hep-th/9802109.}




\baselineskip8pt
\Title{
\vbox
{\baselineskip 6pt{\hbox{ }}{\hbox
{Imperial/TP/97-98/043}}{\hbox{hep-th/9804209}} {\hbox{
  }}} }
{\vbox{\centerline {Einstein spaces in five and seven dimensions}
\vskip4pt
 \centerline      {and non-supersymmetric gauge theories}
}}
\vskip -27 true pt
\centerline  {  Jorge G. Russo{\footnote {$^*$} {e-mail address:
j.russo@ic.ac.uk
 } } 
}

\bigskip

 \centerline {\it  Theoretical Physics Group, Blackett Laboratory,}
\smallskip
\centerline {\it  Imperial College,  London SW7 2BZ, U.K. }

\medskip\bigskip

\centerline {\bf Abstract}
\medskip
\baselineskip10pt
\noindent

A one-parameter family of new solutions 
representing Einstein spaces in $d=5,7$ is presented, and
used to construct non-supersymmetric backgrounds in type IIB and M-theory that asymptotically
approach $AdS_5\times S^5$ and \adss . 
Upon dimensional reduction, the latter gives a type IIA
solution representing a 4-brane with Ramond-Ramond charge,
which
interpolates between the ``near-horizon" non-extremal D4 brane and a geometry connected by T-duality to a new constant dilaton solution in type IIB. 
We discuss the possibility that M-theory on this space may be related to 
a (0,2) six-dimensional field theory on $S^1\times S^1$, with fermions obeying antiperiodic boundary conditions in both circles.

\medskip
\Date {April 1998}
\noblackbox
\baselineskip 14pt plus 2pt minus 2pt

\newsec{Introduction }

The suggestion that large $N$ gauge theory can be described by a string theory
\refs{\thooft, \polya } has been recently  investigated in ref.~\malda , where  specific connections between certain gauge theories and string theories have been proposed.
One of these connections concerns the equivalence between M-theory compactified on \adss \ (i.e.
a seven-dimensional anti de Sitter space and a 4-sphere)
and the six-dimensional (0,2) conformal field theory studied in
\refs{\ewt , \oldstr }. 
When this latter theory is compactified on a two-torus, one obtains a theory which at 
low-energies looks like $SU(N)$ Yang-Mills theory with $\N=4 $ or $\N=0$
supersymmetry, according to the choice of spin structure.
This fact was exploited in a recent paper by Witten \edw , who investigated
a description of the large $N$, $\N=0$  $SU(N)$ Yang-Mills theory in terms of supergravity on
an Einstein space related to the non-extremal M5-brane metric. 
Other proposals to study  large $N$ $\N=0 $ four-dimensional Yang-Mills theory
using supergravity  were investigated e.g. in refs.~\refs{\kach \vaffa -\oduff }. 

A version of the supergravity/CFT correspondence proposes 
that conformal field theory on a manifold $M$ should be described by summing over contributions of Einstein manifolds that have
boundary $M$ at infinity \wittn . It is therefore of interest to
identify possible Einstein manifolds which may be of relevance to a given conformal field theory.
The non-supersymmetric M-theory solution investigated in this work
is given by
$$
ds^2_{11}= f^{-1/3}(r)\left[ 
\( 1-{\mu\ov r^3}\) ^{\a _2}d\tau _2 ^2+ 
 \( 1-{\mu\ov r^3}\)^{\a _1} d\tau_1^2
+   
{ \sum_{i=1}^4 dx_i^2 \ov  \( 1-{\mu\ov r^3}\) ^{\four (\a_1+\a_2 -1) } }\ \right]
$$
\eqn\sori{
\ \ \ \ \ \ \ \ \ \ \ \ +\ f^{2/3}(r) \left[{  dr^2\ov \( 1-{\mu \ov r^3}\)}+
r^2 d\Omega_4^2
\right]\ , \ \ \ \ \ \ \  \mu=r^3_+ -r^3_-\ ,\ \   r_+^3r_-^3=\pi^2N^2l_P^6\ ,
}
$$
dA_3=6\pi N \Omega _4\ ,\ \ \ \  f(r)=1+{r_-^3\ov r^3}\ ,\ \ \ \ \ 
\a_1={1\ov 5} \( 1-\a_2 + 2 \sqrt{4 +2\a_2-6\a_2^2 } \)\ ,
$$
where $\Omega_4$ is the volume form of the unit 4-sphere.
Generically, the geometry contains a naked 
singularity at $r=\mu ^{1/3}$, which we will assume is regularized by $\a' $ corrections.
The (asymptotically flat) solution \sori\ constitutes a one-parameter generalization of the non-extremal M5-brane (the non-extremal M5 brane is the unique solution 
of given charge and mass with $SO(5)$ isometry,  translational and rotational isometries on the brane). 
In the special ``decoupling" limit of \malda \ (one redefines $r\to r^2\ l_P^{3} \ ,\ \ \mu\to\mu \ l_P^{9}$, and takes the limit
$l_P\to 0$ with $r,\mu $ fixed, a factor  $l_P^2$ remaining as 
an overall scale of the metric),
 the metric \sori\  reduces to ($l_P=1$)
$$
ds^2_{11}= r^2 
\( 1-{1\ov r^6}\) ^{\a _2}d\tau _2 ^2+ 
r^2 \( 1-{1\ov r^6}\)^{\a _1} d\tau_1^2
+   
{r^2 \sum_{i=1}^4 dx_i^2 \ov  \( 1-{1\ov r^6}\) ^{\four (\a_1+\a_2 -1) } }\ 
$$
\eqn\hori{
+\ {r^2_0\  dr^2\ov r^2\( 1-{1 \ov r^6}\)}+
\four r^2_0 d\Omega_4^2
\ ,\ \ \ \ \ \ \ \ r_0=2 (\pi N)^{1/3}\ ,
}
where $\mu $ has been scaled away. This is of the form $X\times S^4$, where $X$ is an Einstein space that approaches 
\ads \ in the asymptotic region of large $r$. 
More general Einstein spaces in $d=5$ and $d=7$ will be described at the end of  section 2.

\newsec{Einstein manifolds in five and seven dimensions}

The Anti-de Sitter metric in $d=7$ is given by
\eqn\siete{
ds^2=r^2 \sum_{i=1}^6 dx_i^2 + r_0^2 {dr^2\ov r^2}\ .\ 
}
This obeys
\eqn\eins{
R_{\mu\nu}-\ha g_{\mu\nu} R={15\ov r_0^2}\ g_{\mu\nu}\ .
}
Another Einstein manifold which asymptotically
approaches \ads\ is described by the metric \edw 
\eqn\witt{
ds^2=r^2 \( 1-{1\ov r^6}\) d\tau _2 ^2+ 
{r_0^2 dr^2\ov r^2\( 1-{1\ov r^6}\)}+r^2 d\tau_1^2+r^2 \sum_{i=1}^4 dx_i^2 \ .\ 
}
It can also be
obtained as a special limit of the non-extremal M5 brane,
where $\tau_2 $ plays the role of euclidean time variable.
The metric has a conical singularity at $r=1 $, which can be removed by setting the period of $\tau_2$ equal to $2\pi r_0/3 $.
There are no Killing spinors.

In ref.~\edw\ the space \witt\ was used to
provide a supergravity description for 
the low-energy regime of  four-dimensional  $\N=0 $ super Yang-Mills theory.
One starts with the six-dimensional (0,2) theory on the
circles $C_1,C_2$, described by $\tau_1, \tau_2 $, with  fermions
obeying periodic boundary conditions around the circle  $C_1$, and antiperiodic boundary conditions
 around $C_2$.
This breaks the $\N =4 $ supersymmetry to $\N=0 $.
The construction of the corresponding supergravity description
is made by analogy to the finite temperature gauge theory, where
fermions obey  antiperiodic boundary conditions on the euclidean time direction.
Since this is described by M-theory on the space
 \witt , with $\tau_2 $ playing the role of  euclidean time, one can similarly use \witt\ for the gauge theory with antiperiodic fermions on $C_2$ with $\tau_2$ being the angular variable.
The 4d Yang-Mills gauge coupling will be given by $g_4^2=R_1/R_2$,
where $R_1,R_2$ are the respective radii of $C_1,C_2$
(a discussion of antiperiodic fermions in the case of $AdS_3$ is given in \mastro ).

On $T^2$ there are four spin structures: $(++),\ (+-),\ (-+),\ (--)$,
according to whether the fermions obey periodic $(+)$ or antiperiodic $(-)$ boundary conditions on
the circles $(C_1C_2)$. The $\N=4 $ super Yang-Mills theory corresponding to
$(++)$ is expected to be described by the M-theory background constructed with \siete\ and a 4-sphere \malda .
According to \edw , the $\N=0 $ Yang-Mills theory corresponding to $(+-)$ should be described 
in terms of the space \witt . Similarly, the $(-+)$ case should be described by 
the same space changing $\tau_1 \leftrightarrow \tau_2 $.
It is not obvious whether there is a metric that can be used
to describe the $\N=0 $ Yang-Mills theory $(--)$ with antiperiodic
fermions in both circles, $C_1$ and $C_2$.

The metric \witt\ explicitly breaks the $SL(2,{\bf Z})$ 
symmetry of the 2-torus because of the different treatment of
$C_1$ and $C_2$.  In the case $(--)$, just as in the $(++)$ case \siete ,
one expects that the appropriate metric will be symmetric under
$\tau_1 \leftrightarrow \tau_2 $.
Thus one should look for Einstein spaces with metrics of the form
\eqn\forma{
ds^2=f_1(r)(d\tau_1^2 + d\tau_2^2 )+f_2(r)dr^2+f_3(r) \sum_{i=1}^4 dx_i^2 \ .\  
}
It is easy to show that there are only three solutions to the Einstein equations \eins\
of the form \forma \ (modulo locally equivalent geometries). 
One of the functions in \forma , say $f_3$, can be a absorbed into a redefinition of $r$.
The ``33" component of the Einstein equations then determines
$f_2$ in terms of derivatives of $f_1$. The remaining components
give a differential equation for $f_1$, which is solved by the 
$AdS_7$ space \siete\ and by
\eqn\zzz{
ds^2=r^2 \( 1-{1\ov r^6}\) ^\a (d\tau _2 ^2+ d\tau_1^2)
+   {r_0^2 dr^2\ov r^2\( 1-{1\ov r^6}\)}+
{ r^2\ov  \( 1-{1\ov r^6}\) ^{\ha \a -\four } } \sum_{i=1}^4 dx_i^2  \ ,\ 
}
$$
\a =\sei (1+\sqrt{10})\ .
$$
The third solution  is similar to \zzz\ with a change of sign in front of $1/r^6$.

By the redefinition  $\td r^6=r^6-1$, the metric \zzz\ can be put into the form
\eqn\xxx{
ds^2={\td r^2\ov \( 1+{1\ov \td r^6}\) ^\b } (d\tau _2 ^2+ d\tau_1^2)
+   {r_0^2 d\td r^2\ov \td r^2\( 1+{1\ov \td r^6}\)}+
 \td r^2  \( 1+{1\ov \td r^6}\) ^{\ha \b + \four } \sum_{i=1}^4 dx_i^2  \ ,\ 
}
$$
\b =\sei (-1+\sqrt{10}) \ .
$$

Let us also present the one-parameter family of solutions that interpolates between \zzz\ and \witt . It is given by
\eqn\yyy{
ds^2=r^2 \( 1-{1\ov r^6}\) ^{\a _2}d\tau _2 ^2+ 
r^2 \( 1-{1\ov r^6}\)^{\a _1} d\tau_1^2
+   {r_0^2 dr^2\ov r^2\( 1-{1\ov r^6}\)}+
{ r^2\sum_{i=1}^4 dx_i^2 \ov  \( 1-{1\ov r^6}\) ^{\four (\a_1+\a_2 -1) } }\ ,\ 
}
with $\a_1$ and $\a_2$ related by
\eqn\nnb{
\a _1= {1\ov 5} \( 1-\a_2 \pm 2 \sqrt{4 +2\a_2-6\a_2^2 } \)\ ,
}
or, in terms of $\a_\pm =\a_1\pm \a_2 $,
\eqn\nnz{
\a _+= {1\ov 3} \( 1  \pm  \sqrt{10 -6\a_-^2 } \)\ .
}
The two possible signs in \nnb\ correspond to the choice of sign
in front of $1/r^6$ (combined with a redefinition of $r$).
As the ``near-horizon" non-extremal M5 brane \witt , this metric breaks all the supersymmetries, and asymptotically
approaches the \ads\ metric \siete .
The geometries \zzz\ and \yyy\ exhibit a curvature singularity at $r=1$ (while $R$ and $(R_{\mu\nu})^2$ are constants, $(R_{\mu\nu\rho\sigma })^2$ diverges at this point).

More general solutions --~which do not have the full Poincar\' e symmetry in the four-dimensional space $x_i$~-- can also be constructed. In particular, the following one is a two-parameter family of Einstein spaces which includes \yyy :
\eqn\yyyg{
ds^2=r^2 \bigg[ h ^{\a _1}d\tau _1 ^2+ h ^{\a _2}d\tau _2 ^2+
h ^{\a _3}d\tau _3 ^2+
{\sum_{i=1}^3 dx_i^2 \ov  h^{{1\ov 3}(\a_1+\a_2+\a_3 -1)} }\bigg] 
+ {r_0^2 \ dr^2\ov r^2\( 1-{1\ov r^6}\)}\ ,
}
$$
h(r)=1-{1\ov r^6}\ ,
$$
with $\a_1, \a_2,\a_3$  related by
$$
2\a_1^2+2\a_2^2+2\a_3^2+\a_1\a_2+\a_2\a_3+\a_1\a_3-
\a_1-\a_2-\a_3=1\ .
$$

Einstein spaces in $d=5$ that asymptotically approach $AdS_5$
can be similarly found. The one-parameter family analogous to
\yyy\ is given by
$$
ds^2_5= r^2 \( 1-{1\ov r^4}\) ^{\a _2}d\tau _2 ^2+ 
r^2 \( 1-{1\ov r^4}\)^{\a _1} d\tau_1^2
+   {r_0^2 dr^2\ov r^2\( 1-{1\ov r^4}\)}
$$
\eqn\dddc{
+\ 
 r^2\( 1-{1\ov r^4}\) ^{\ha (1-\a_1-\a_2 ) }   (dx_1^2+dx_2^2)
\ ,\ 
}
with $\a_1$ and $\a_2$ related by
\eqn\nnb{
\a _1= {1\ov 3} \( 1-\a_2 \pm 2 \sqrt{1 +\a_2-2\a_2^2 } \)\ .
}
Special cases are $\a_1=1,\a_2=0$ (related to the non-extremal D3-brane)
and $\a_1=\a_2=\four (1+\sqrt{3})$.

\newsec{M-theory and type II solutions}

The Einstein spaces \zzz\ and \yyy\ (or the more general one \yyyg ) can be used to construct new
solutions of eleven-dimensional supergravity by adding
a 4-sphere and flux of the 4-form field strength. 
In particular, 
the M-theory background corresponding to \zzz\ is given by 
\eqn\mzz{
ds^2_{11}=r^2 \( 1-{1\ov r^6}\) ^\a (a_2^2 d\t _2 ^2+ a_1^2 d\t_1^2)
+ {r_0^2 \ dr^2\ov r^2\( 1-{1\ov r^6}\)}+
{ r^2\sum_{i=1}^4 dx_i^2 \ov  \( 1-{1\ov r^6}\) ^{{\a\ov 2} -\four } }+
\four r_0^2 d\Omega_4^2 \ ,\ 
}
$$
dA_3=6\pi N \Omega _4\ ,\ \ \ \ \ \ r_0=2l_P(\pi N)^{1/3}\ ,
$$
where we have set $\tau_1=a_1\theta_1 $, $\tau_2=a_2\theta_2 $, $\t_{1,2} =\t_{1,2}+2\pi $.
By dimensional reduction along
$\t _1 $, one finds the following
type IIA solution:
$$
ds^2_{\rm IIA} = a_1a_2^2 r^3 \( 1-{1\ov r^6}\) ^{ {3\ov 2}\a } d\t_2^2
+{a_1r_0^2 \ dr^2\ov r \( 1-{1\ov r^6}\)^{1-{\a \ov 2}} } +
a_1 r^3 \( 1-{1\ov r^6}\) ^{\four } \sum_{i=1}^4 dx_i^2 
$$
\eqn\ddaa{
+\ 
\four a_1 r_0^2 r \( 1-{1\ov r^6}\) ^{{\a\ov 2} } d\Omega_4^2 \ \ ,\ 
}
\eqn\dila{
e^{ {2\ov 3}\phi_A }= a_1 r \( 1-{1\ov r^6}\) ^{{ \a\ov 2} } \ .
}
By T-duality along $\t_2$ we obtain the type IIB solution:
$$
ds^2_{\rm IIB} = {1\ov a_1a_2^2 r^{3} } \( 1-{1\ov r^6}\) ^{-{3\ov 2}\a } d\td \t_2^2
+{a_1r_0^2 \ dr^2\ov r \( 1-{1\ov r^6}\)^{1- {\a\ov 2} } }
+\  a_1 r^3 \( 1-{1\ov r^6}\) ^{\four } \sum_{i=1}^4 dx_i^2 
$$
\eqn\ddbb{
+\four a_1 r_0^2 r \( 1-{1\ov r^6}\) ^{\ha \a } d\Omega_4^2 \ ,\ 
}
\eqn\bdila{
e^{\phi_B }= {a_1\ov a_2}\ .
}
Noteworthy, this is a solution with constant dilaton (this can be anticipated from the form of the eleven-dimensional metric \mzz ). 
In the Witten model \witt , the type IIB  string coupling constant diverges 
at  $r=1$ (the type IIA string coupling diverges at large $r$). 
The  Einstein-frame metric corresponding to the type IIB solution \ddbb\ is 
thus given by
$$
ds^2_{E} =(a_1a_2)^{1/2}
 \bigg[ {1\ov (a_1a_2)^{2} r^{3}} \( 1-{1\ov r^6}\) ^{-{3\ov 2}\a } d\td \t_2^2
+{r_0^2\ dr^2\ov r \( 1-{1\ov r^6}\)^{1-{\a \ov 2}} }
$$
\eqn\eebb{
+\ 
r^3 \( 1-{1\ov r^6}\) ^{\four } \sum_{i=1}^4 dx_i^2 +
\four r_0^2 r \( 1-{1\ov r^6}\) ^{{ \a \ov 2}} d\Omega_4^2 \bigg]\ .\ 
}

Likewise, one constructs the  M-theory solution \hori\ 
corresponding to the Einstein manifold \yyy \ times a 4-sphere. 
Upon dimensional reduction, the metric \hori\ gives
$$
ds^2_{\rm IIA} = a_1a_2^2 r^3 \( 1-{1\ov r^6}\) ^{ \a_2+ {\a_1\ov 2} } d\t_2^2 +{a_1r_0^2 \ dr^2\ov r \( 1-{1\ov r^6}\)^{1-{\a _1\ov 2}} } 
$$
\eqn\ddyy{
+\ 
a_1 r^3 \( 1-{1\ov r^6}\) ^{\four (1+\a_1-\a_2 )} \sum_{i=1}^4 dx_i^2 
+ \four a_1 r_0^2 r \( 1-{1\ov r^6}\) ^{{\a_1\ov 2} } d\Omega_4^2 \ ,\ 
}
\eqn\dilan{
e^{ {2\ov 3}\phi_A }= a_1 r \( 1-{1\ov r^6}\) ^{{ \a _1\ov 2} } \ ,
}
where  $a_1, a_2$ have been restored.
This corresponds to the `decoupling' limit of the reduction of \sori,
$$
ds^2_{\rm IIA} = f^{-1/2}(r)
\bigg[ \left( 1-{\mu\ov r^3}\right)
^{ \a_2+{\a_1\ov 2} } d\tau_2^2 +
\left( 1-{\mu\ov r^3}\right)
^{{1\ov 4}(1+\a_1-\a_2)}\sum_{i=1}^4 dx_i^2\bigg]
$$
\eqn\zzyy{
+\ f^{1/2}(r)\left( 1-{\mu\ov r^3}\right)
^{\a_1\ov 2}\bigg[ { dr^2\ov \left( 1-{\mu\ov r^3}\right)}
+r^2 d\Omega_4^2\bigg]\ ,
}
\eqn\zila{
e^{ {2\ov 3}\phi_A }=  f^{-1/6}(r) \left( 1-{\mu\ov r^3}\right)
 ^{{ \a _1\ov 2} } \ ,
}
representing a family of 4-branes with singular horizon. 
This background generalizes
the euclidean non-extremal D4 brane and has a similar structure,
i.e. the same isometries and same RR 3-form (although the coupling vanishes
at the horizon, the Riemann tensor is singular; $\a ' $ corrections
should be important near $r=\mu^{1/3}$).
More general type IIB and M-theory backgrounds based on the 
solutions \yyyg , \dddc\ can be constructed in a similar way.

\newsec{Possible connections with gauge theories}

Given the uniqueness of the solution of the form \forma ,
one would like to investigate the possibility 
that M-theory on
the space \mzz\ may be related to a six-dimensional (0,2)
theory compactified on circles $C_1,C_2$ with
antiperiodic boundary conditions for the fermions in both circles.

In terms of type IIB string theory, one is compactifying on the space \ddbb ;
this corresponds to M-theory on \mzz\ in
the limit $R_1, R_2\to 0$ at fixed $g_4^2\equiv R_1/R_2$, which preserves
the $SL(2,{\bf Z} )$ reparametrization symmetry of the 2-torus. 
If there is a field-theory boundary description for this string model, 
it should therefore be an $\N=0$  $d=4 $ field theory
with coupling constant $g_4^2$,
$S$-duality symmetry group $SL(2,{\bf Z} )$ and a global $SO(5)$ symmetry (although  supersymmetry is broken, 
the $S$-duality symmetry under $g_4^2 \to 1/g_4^2$ should still hold, since on the supergravity side there is no distinction
between the two directions $\tau_1$ and $\tau _2$).
Since the corresponding type IIA string model  
has  the (near-horizon) $N$ D4-brane background as asymptotic limit,  
the field theory in question  presumably describes
the low-energy regime of an $\N=0$   $SU(N)$
$d=4+1 $ super Yang-Mills compactified on  $S^1$.
 Because of the supersymmetry breaking boundary conditions,
fermions and scalar particles are expected to get a mass. As a result, the low-energy gauge theory should be pure $SU(N)$ Yang-Mills theory.

The solution \mzz\ is valid
for $r_0\gg l_P$ (i.e. $N\gg 1$), and in the region $r\gg 1 $.
If $R_1,\ R_2$ denote the radii of two circles $C_1,\ C_2$ measured
at $r=\infty $, then $a_1$ and $a_2$ are related by
$
{a_1\ov a_2}={R_1\ov R_2}=g_4^2\ .$
One is interested in the large $N$ limit
with $g_4\to 0$ and  fixed $ g_4^2 N$, i.e. fixed
${Na_1\ov a_2}\ .$
To understand the large $N$ limit on the supergravity side
(in particular, properties of the spectrum \refs{\gkp,\hoog ,\wittn}),
it is necessary to determine how $a_1a_2$ depends
on $N$ (cf. eq.~\eebb ).
In the case of the Witten model \witt , the value of $a_2$ was fixed by demanding
the metric to be free of conical singularity at $r=1$.
In the present case, the solution does not apply near $r=1$;
the problem seems sensitive to $\a' $ corrections.
 As long as the singularity at $r=1$ is regularized by $\a ' $ corrections,
a mass gap is expected, as  can be proved in a similar way as in the
AdS case: there are no oscillatory solutions to the Laplace equation
at infinity, so the spectrum of normalizable solutions
that are regular at $r=1$  must therefore be discrete {\edw }.

It is important that, 
despite the naked singularity, the euclidean action
$$
I=- {1\ov 16\pi G} \int d^7x \ \sqrt{g}\ \big(R+ {15\ov  r_0^{2} } \big) \ ,
$$
is ultraviolet finite, since the scalar curvature is finite, $R=-21/r_0^2$,
so that $I=6(8\pi G)^{-1}\int d^7x\ \sqrt{g}$. By regularizing the volume
by making $x_i$ periodic and $r\in (1,L)$,
it is easy to see (along  similar lines as in~\edw ) that $\lim _{L\to\infty }\big[I_0(L)-I(L)\big]=0$, where
$I_0,\ I$ are the actions corresponding to \witt\ and \yyy , respectively.

A natural question is whether there is a  field-theory boundary
description for M-theory on the more general background \hori .
This solution interpolates between the Witten model \witt\ 
(describing gauge theory with periodic fermions
on $C_1$) and the model \mzz\ (which is perhaps relevant to 
the gauge theory with antiperiodic fermions on $C_1,C_2$). 
This suggests that M-theory on \hori \ 
may describe Yang-Mills theory 
with a magnetic field configuration in an internal direction, since magnetic field
configurations can interpolate between
periodic and antiperiodic boundary conditions for the fermions \melvin . For small $\a_1 $, the model \yyy\ can be regarded
as a deformation of the Witten model \witt ; it is conceivable that such deformation  corresponds to adding certain perturbations to the Yang-Mills theory at the boundary.

Interestingly, a contingent description of M-theory on the space \mzz\ in terms of a gauge theory at the boundary
would also provide a setting for an understanding
of naked singularities in string theory, in the spirit of ref.~\horro .

\bigskip
\bigskip

\noindent{\bf Acknowledgements}
\medskip
The author wishes to thank A.~Tseytlin for comments.
  This work is supported by the European
Commission TMR programme grant  ERBFMBI-CT96-0982.

\vfill\eject
\listrefs

\bye